\documentclass[aps,reprint,superscriptaddress]{revtex4-1}

\usepackage{amsmath}    
\usepackage{graphicx}   
\usepackage{color}      
\usepackage{subfigure}  
\usepackage{hyperref}   
\usepackage{todonotes}
\usepackage[latin1]{inputenc}

\begin{document}

\title{Electronic properties and morphology of Cu-Phthalocyanine - C$_{60}$ composite mixtures}
\author{Friedrich Roth}
\affiliation{Center for Free-Electron Laser Science / DESY, Notkestra\ss e 85, D-22607 Hamburg, Germany}
\author{Cosmin Lupulescu}
\affiliation{Inst. of Optics and Atomic Physics, TU Berlin, Stra\ss e des 17. Juni 135, D-10623 Berlin, Germany}
\author{Tiberiu Arion}
\affiliation{Center for Free-Electron Laser Science / DESY, Notkestra\ss e 85, D-22607 Hamburg, Germany}
\affiliation{Institut f\"ur Experimentalphysik, Universit\"at Hamburg, Luruper Chaussee 149, D-22761 Hamburg, Germany}
\author{Erik Darlatt}
\affiliation{Physikalisch-Technische Bundesanstalt (PTB), Abbestra\ss e 2-12, D-10587 Berlin, Germany}
\author{Alexander Gottwald}
\affiliation{Physikalisch-Technische Bundesanstalt (PTB), Abbestra\ss e 2-12, D-10587 Berlin, Germany}
\author{Wolfgang Eberhardt}
\affiliation{Center for Free-Electron Laser Science / DESY, Notkestra\ss e 85, D-22607 Hamburg, Germany}
\affiliation{Inst. of Optics and Atomic Physics, TU Berlin, Stra\ss e des 17. Juni 135, D-10623 Berlin, Germany}
\date{\today}

\begin{abstract}
Phthalocyanines in combination with C$_{60}$ are benchmark materials for organic solar cells. Here we have studied the morphology and electronic properties of co-deposited mixtures (blends) of these materials forming a bulk heterojunction as a function of the concentration of the two constituents. For a concentration of 1:1 of CuPc:C$_{60}$ a phase separation into about 100\,nm size domains is observed, which results in electronic properties similar to layered systems. For low C$_{60}$ concentrations (10:1 CuPc:C$_{60}$) the morphology, as indicated by Low-Energy Electron Microscopy (LEEM) images, suggests a growth mode characterized by (amorphous) domains of CuPC, whereby the domain boundaries are decorated with C$_{60}$. Despite of these markedly different growth modes, the electronic properties of the heterojunction films are essentially unchanged.
\end{abstract}

\maketitle

\section{Introduction}

Organic electronics offer the promise of easily producible, low cost mechanically flexible electronics for mass production applications. Among these are logic circuits, organic light emitters, and photovoltaics. While there has been considerable commercial success with respect to low speed logic circuits \cite{Crone2000,Derycke2001,Credi1997,Magri2006} and organic LED's,\cite{Friend1999,Meerheim2009,Weichsel2012} organic photovoltaics are still somewhat elusive because of generally low efficiencies and less durability. \cite{Leo2008} Important progress in the design of organic solar cells covering both new functional materials as well as theoretical developments has been recently achieved by employing x-ray spectroscopy as a fundamental characterization tool for organic materials.\cite{Himpsel2012,Pickup2013} 


C.\,W. Tang \cite{Tang1986} was the first to produce an organic solar cell based upon a bilayer system consisting of two adjacent layers of organic compounds with different electronic functionality. The function of one layer, which incidentally was Cu-Phthalocyanine (CuPC), was to serve as light absorber, while the second one, a perylene derivative served as electron acceptor. At that time this cell surpassed by far any other organic PV cell in terms of light conversion efficiency (1\%) and performance (fill factor). In terms of functionality the interface between these two layers resembles the p-n-junction which in a conventional semiconductor is established by doping. In detail however the explanation for the superior performance is a bit more complex. Light absorption mostly generates excitons in the PC and the interface provides an efficient pathway for dissociation of the excitons into separate electronic charges. Thus recombination losses are suppressed and the open circuit voltage is largely determined by the electronic properties of the interface and not by the contact layers.

A substantial new development was the discovery of Sariciftci and Heeger \cite{Sariciftci1992} that adding C$_{60}$ to an organic conductor  improves the photoconductivity by orders of magnitude. The model proposed was that C$_{60}$, because of it's high electron affinity, \cite{Gunnarsson1995} is very efficient in exciton dissociation, separating the charges initially created in the photoabsorption process and thus preventing recombination. Obviously, if this improves the photoconductivity it should have similar consequences for the efficiency of organics in photovoltaic applications.

This model was substantiated by photoemission studies of the band alignment in thin film systems consisting of C$_{60}$ and various phthalocyanines.\cite{Schlebusch1996,Schlebusch1999,Kessler1998} These studies established that the alignment of the topmost occupied bands of C$_{60}$ films and various phthalocyanine layers in contact is indeed such, that the excitonic or polaronic state created by photoabsorption in the chromophore (PC) is higher in energy than a state, where the electron is transferred into the LUMO of C$_{60}$ and the hole remains at the PC. Since then there were many studies characterizing the electronic properties of layered systems of C$_{60}$ and organic molecular films, especially PC's, under various processing conditions published in the literature. The current state of knowledge about these systems and their role in organic photovoltaic devices has been summarized recently. \cite{Fahlman2013,Opitz2012}

In these layered systems or so called planar heterojunctions the performance optimization limits the layer thickness to the exciton diffusion length, which is on the order of 10\,nm. For an efficient light absorption however a film thickness of about an order of magnitude larger (100\,nm) is desired. Consequently the attention has been transferred to mixed blends of organic molecules and C$_{60}$ for such devices. For a review of the design criteria we refer to the paper of Heremans $et\,al.$,\cite{Heremans2009} while the general status of the development of organic photovoltaics has been reviewed by Leo and coworkers.\cite{Leo2008} From previous experiments it was found that a well mixed phase separated blend exhibits the best device performance.\cite{Schuenemann2012} This means that small crystalline domains, especially of C$_{60}$, well embedded into a PC matrix is the preferred configuration for bulk heterojunction solar cell applications. The individual regions/domains should be sufficiently small to enable efficient exciton diffusion and separation, while the crystallinity improves the charge carrier mobility, at least for the electrons within the C$_{60}$ crystallite. However from these studies a very puzzling question remains and that is why the performance varies only very little over quite large changes in concentration of C$_{60}$ in the blends. Especially the exciton dissociation remains almost unchanged as a function of the C$_{60}$ concentration.\cite{Schuenemann2012}

While there have been Grazing Incidence X-Ray Absorption Spectroscopy (GIXAS) and Atomic force microscopy (AFM) measurements correlated with the device performance under different preparation conditions,\cite{Schuenemann2012} we present here an electron spectroscopy investigation of the electronic properties of CuPC:C$_{60}$ blends in connection with their morphology. There have been observations that the morphology changes by annealing, deposition temperature, or substrate conditions,\cite{Schuenemann2012,Kim2011} we here show that solely a change in the relative composition ratio of the blend, even while keeping the sample at room temperature, will result in a different morphology of the bulk heterojunction. Despite of this change in morphology, the electronic properties are quite similar. The electronic properties are probed by photoemission, while the morphology is studied by Low-Electron Enery Microscopy (LEEM), which results in an image of the local electronic properties with a spatial resolution better than 5\,nm.

\section{Experimental}

\begin{figure}[h]
\includegraphics[width=0.9\linewidth]{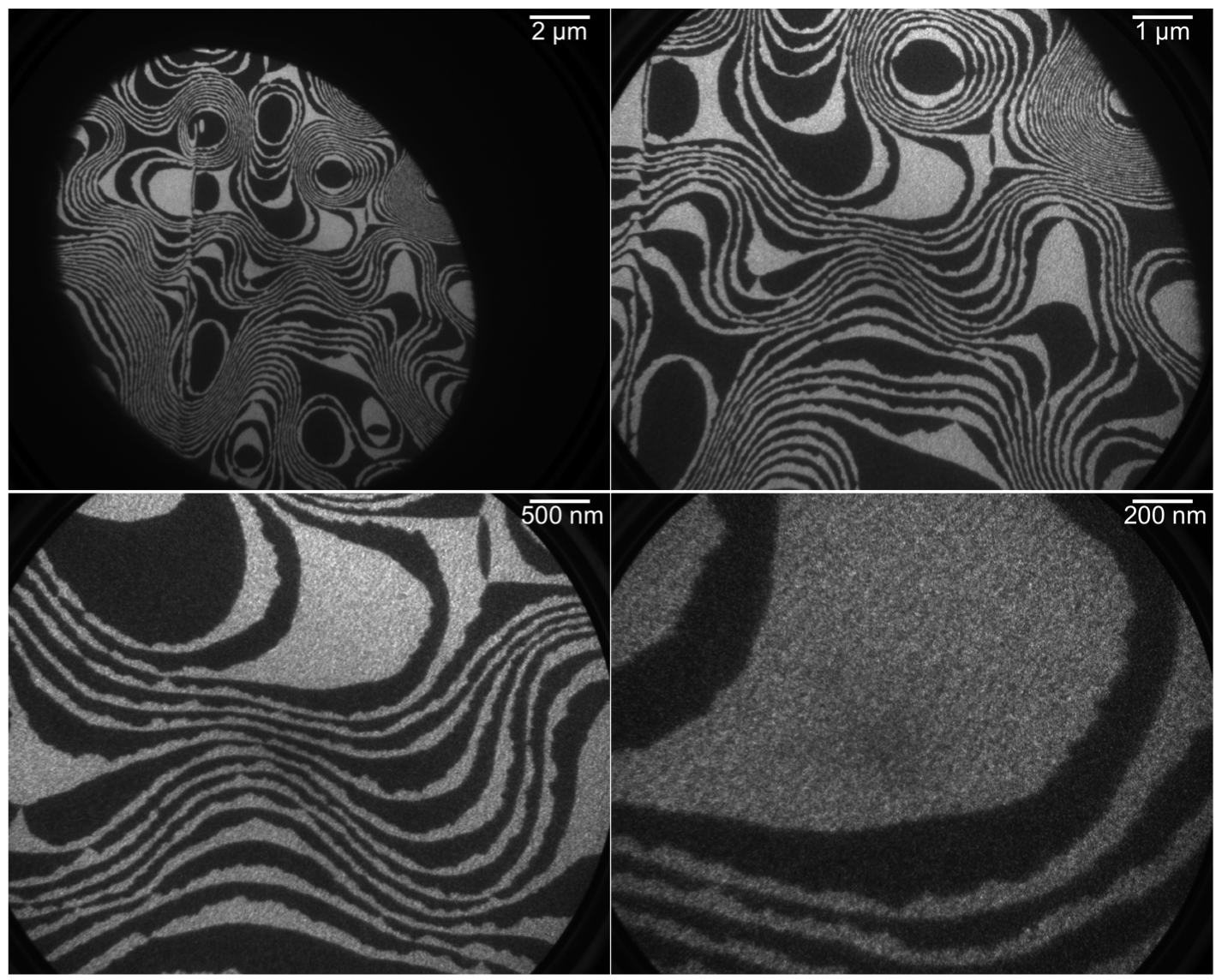}
\caption{LEEM images of a clean Si(100) wafer surface. The complete field of view of the microscope was decreased in the four individual images from 20\,$\mu$m (upper left image) to 2\,$\mu$m (lower right image), as suggested by the respective scale bars.}
\label{f1}
\end{figure}

We have chosen a Si(100) 2x1 reconstructed wafer as substrate in order to have well characterized reproducible substrate conditions, where the electronic properties are well established and the surface morphology is characterized by the 2x1 surface reconstruction resulting in stripes of typically \textgreater\,100\,nm width with atomically sharp interfaces in two orthogonal orientations (a typical image is shown in Fig.\,\ref{f1}). The fullerene and phthlalocyanine films were grown on such substrates by $in\,situ$ (co)evaporation of C$_{60}$ and CuPC from two spatially separated effusion cells to obtain a 7\,nm thick film of pristine substances and 14\,nm of diluted species, respectively. The growth at room temperature was monitored by a quartz crystal microbalance to ensure homogeneous and continuous films over the 5x5\,mm$^2$ Si(100) wafer. The samples were prepared in the preparation chamber (base pressure of  $<$\,2\,x\,10$^{-8}$\,mbar) connected to the analysis chamber of the iDEEAA apparatus (base pressure of  $<$\,5\,x\,10$^{-10}$\,mbar).\cite{Lupulescu2013}

Photoelectron spectra were recorded with the SCIENTA R4000 hemispherical electron spectrometer in the iDEEAA end station.\cite{Lupulescu2013} The instrument was installed at Berlin's newest synchrotron radiation facility: Metrology Light Source (MLS), located in the Willy-Wien-Laboratorium of the Physikalisch-Technische Bundesanstalt (PTB).\cite{Gottwald2012} Whereas the BESSY II electron storage ring is a multi-user site, the MLS can operate under various, stable ring conditions, attuned to user-specified individual measurement tasks. For this work, the iDEEAA end station was installed at the Insertion Device Beamline (IDB) whereby a 30.5 period undulator of 125\,mm length (U125) serves as insertion device. The monochromator combines normal incidence (NI) with grazing incidence (GI) geometries and allows for monochromatic radiation from approximately 1.5\,eV to 10\,eV in the NI mode and 10\,eV to 280\,eV in the GI mode. The flux is about of 10$^{12}$\,photons/s at a 100\,mA ring current and the resolving power is better than 700 (GI mode) and 2000 (NI mode). The spot size on the sample is about 1.7\,mm horizontal and 0.1\,mm vertical. At a pass energy of 20\,eV the electron spectrometer was operated at an energy resolution of 15\,meV.

\begin{figure*}
\includegraphics[width=0.9\linewidth]{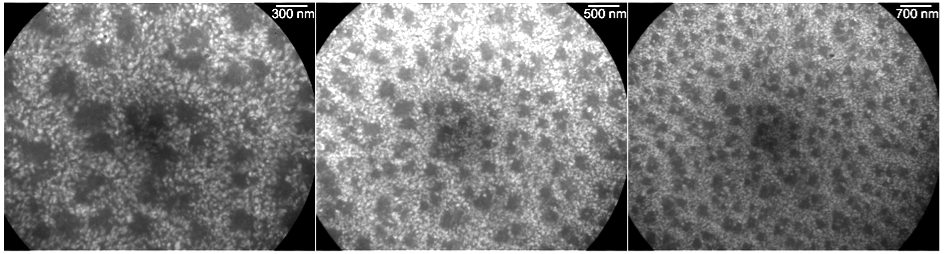}
\caption{LEEM images of the surface of a CuPC/C$_{60}$ mixture deposited on a Si wafer substrate. The mixing ratio for preparing the sample ($in\,situ$) was CuPC:C$_{60}$ = 1:1, and the exposure time for the measurement was 5\,s.}
\label{f2}
\end{figure*}

\begin{figure*}
\includegraphics[width=0.9\linewidth]{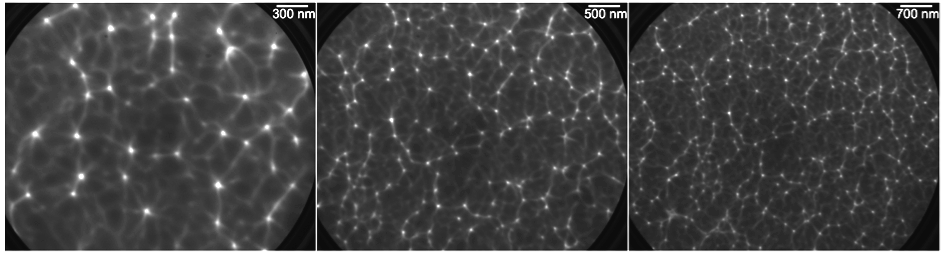}
\caption{LEEM images of the surface of a CuPC/C$_{60}$ mixture deposited on a Si wafer substrate. The mixing ratio for preparing the sample ($in\,situ$) was CuPC:C$_{60}$ = 10:1, and the exposure time for the measurement was 5\,s.}
\label{f3}
\end{figure*}

The sample morphology was investigated using a commercial Low-Electron Energy Microscope (LEEM)/Photoelectron Emission Microscope (PEEM) instrument (SPECS FE-LEEM P90) by following the same sample preparation procedure as for the photoelectron spectroscopy measurements. The effusion cells were mounted inside the load-lock chamber of the LEEM/PEEM instrument and fresh samples were prepared at same conditions as in the photoemission experiment. In our LEEM/PEEM instrument, a cold field emission gun, situated at the top end emits an electron beam with an energy of 15\,keV and about 250\,meV bandwidth.\cite{Tromp2010} The sample illumination is controlled by a combination of a magnetic gun lens and a condenser lens, which also focuses the electron beam towards the entrance plane of a magnetic prism array. Situated at the center of the instrument, the magnetic prism array spatially separates the condenser column from the projector column. It also deflects the electron beam by 90$^\circ$ from the gun into the objective lens system towards the sample. Following the reflection from the sample, the beam is re-accelerated to 15\,keV and re-enters the objective lens. On a crystalline sample, the low energy electrons undergo diffraction and a low energy electron diffraction (LEED) pattern is formed in the backfocal plane of the objective lens, which also forms a real-space image of the sample. In the PEEM modus, the sample is illuminated with UV photons from a mercury-vapor lamp, generating photoelectrons. They are accelerated into the objective lens, forming again a real-space image. The photoelectrons are emitted with a certain angular distribution that can be observed in the objective lens backfocal plane. An energy filter is added to the instrument in the form of a single scanning aperture slit to the objective transfer optics that can be used to obtain energy filtered images, energy filtered photoelectron momentum distributions, as well as full energy resolved E-$k$ data.\cite{Tromp2009} 

The objective transfer lens refocuses the LEED pattern in the entrance plane of the magnetic prism array. A further magnified real-space image is formed on the diagonal plane of the magnetic prism array. The electron beam is then deflected downwards to the projector lens column by the prism array at 90$^\circ$ angle. A contrast aperture is used to select the diffracted beam for image formation. Both bright field and dark field imaging conditions are therefore possible. The sample image and the diffraction pattern are inspected onto an image-viewing screen situated after a multichannel plate (MCP). 

\section{Results and discussion}

The morphology of the films was investigated by using LEEM. To demonstrate the power of this technology for characterizing our thin films, we imaged in LEEM mode the surface of the Si(100) crystal, prior to the deposition of the organic films. This is shown in Fig.\,\ref{f1}, where the images exhibit the domain pattern of the (2x1) reconstruction of this surface with atomically sharp boundaries. From these images, a spatial resolution of better than 5\,nm can be derived.

The LEEM images of the different blends of CuPC:C$_{60}$ after deposition onto these reconstructed Si surfaces are shown in Figs.\,\ref{f2} and \ref{f3}.  First of all, the substrate reconstruction is not visible anymore. However the blends exhibit distinct differences in the morphology. For the 1:1 mixed film, shown in Fig.\,\ref{f2}, a phase separation into domains of several tens of nm size is observed. Interestingly, for mixed films of 1:1 composition a phase separation into domains of amorphous CuPC and crystalline C$_{60}$ has been observed previously using GIXAS.\cite{Schuenemann2012} 

For significantly smaller admixtures of C$_{60}$, no crystalline C$_{60}$ domains were found.\cite{Schuenemann2012} This is perfectly consistent with our real space images shown here for both compositions. The images of the low C$_{60}$ content blend (10:1 CuPC:C$_{60}$) are shown in Fig.\,\ref{f3} indicating a quite different morphology. Here the image is characterized by many bright spots connected by some faint lines. Considering that the contrast in these images is derived from electron scattering, these images are suggestive of the formation of a network of conductive pathways within the film. While this seems speculative at present, it is consistent with a growth mode of CuPC in domains while the domain boundaries are decorated by C$_{60}$. This conveniently also explains the relatively unchanged performance concerning exciton separation in mixed films of low C$_{60}$ content. Regions of C$_{60}$ are accessible within the exciton diffusion length, even for low concentration blends. The higher electron mobility in the crystalline C$_{60}$ regions of the phase separated system only has a marginal influence on device performance. \cite{Schuenemann2012}

The growth of pure CuPC on various cleaned Si surfaces has also been studied by X-ray scattering.\cite{Kim2011} Here  typically 10\,nm size domains of nanocrystalline CuPC, standing edge on tilted in various orientations on the surface, were found, separated by small amorphous regions. This growth mode is also suggestive of the fact, that for the low concentration C$_{60}$ blend, the C$_{60}$ preferentially is found in the regions between the nanocrystalline CuPC.

\par

\begin{figure}
\includegraphics[width=0.7\linewidth]{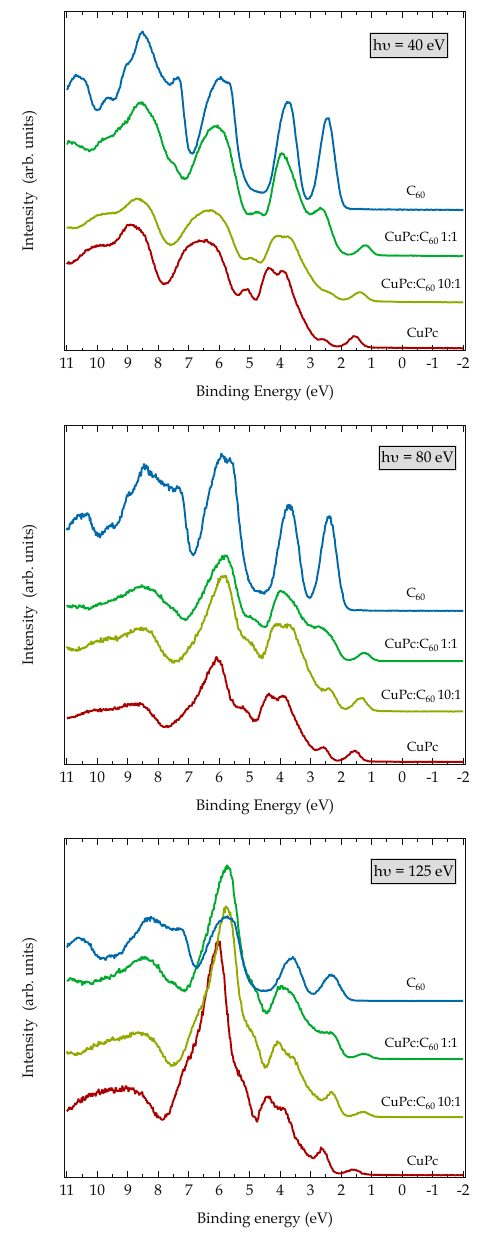}
\caption{Photoemission spectra for pristine CuPC and C$_{60}$ as well as for two mixtures of CuPC:C$_{60}$ (10:1 and 1:1) for various photon energies ($h\nu$ = 40, 80, and 125\,eV). The energy scale is referenced to the Fermi level of the substrate.}
\label{f4}
\end{figure}

How is this remarkable difference in morphology reflected in the electronic properties? This question is addressed in the photoemission spectra shown subsequently. Fig.\,\ref{f4} shows photoemission spectra of nominally 7\,nm (for the pure) and 14\,nm (for the mixtures) thick films taken at three different photon energies. For each photon energy the spectra are shown for the pure CuPC, a 10:1 CuPC:C$_{60}$ film, a 1:1 CuPc:C$_{60}$ film and a pure C$_{60}$ film. The energy scale is referenced to the Fermi level of the substrate. The film thickness was chosen such that the substrate does not visibly contribute to the spectra. We checked for sample charging by varying the incident photon flux density by more than an order of magnitude and for sample deterioration by taking these spectra repeatedly. No charging and/or deterioration of the samples were detected.

\begin{figure*}
\includegraphics[width=0.95\linewidth]{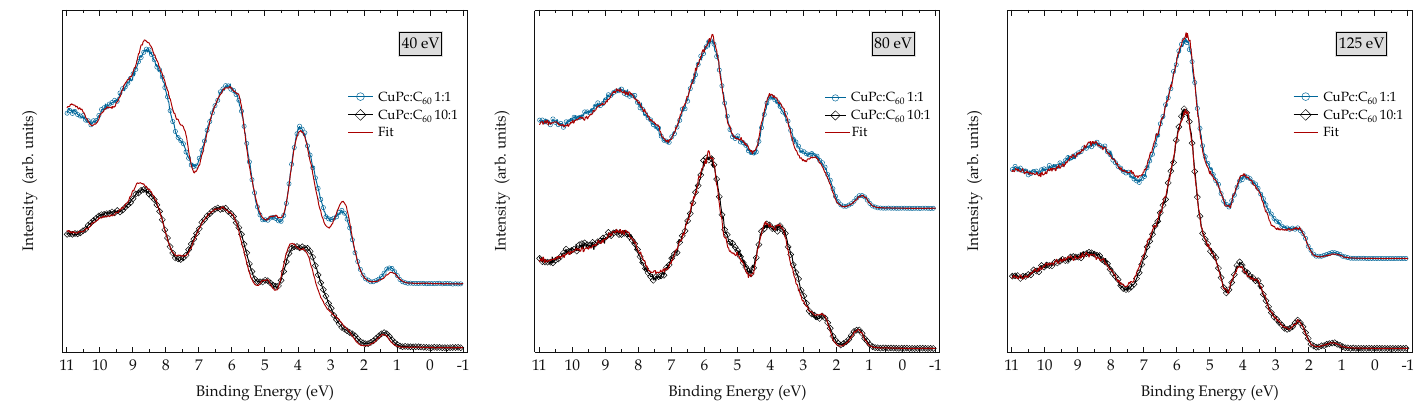}
\caption{Comparision between measured photoemission spectra of the two mixtures of CuPC:C$_{60}$---10:1 (black diamonds) and 1:1 (blue circles)---with results of the fit using the spectra of the pure compounds (for details see text) for three different excitations energies.}
\label{f5}
\end{figure*}

From these spectra it is obvious that the level alignment changes from the pure films to the mixtures/blends. This is very clearly visible when observing the highest energy feature, which is assigned as the emission of the HOMO of CuPc. The position of the C$_{60}$ derived emission is not as easily derived especially for the low concentration (10:1) blends. 

Accordingly we have resorted to a different procedure for analyzing these spectra. We have taken the spectra of the mixed films at various photon energies and fitted these using the spectra of the pure compounds allowing for an energy shift and intensity variation. The result of this four parameter fit, two intensities and two shifts, is shown in Fig.\,\ref{f5} as red lines for three different photon energies (40, 80, and 125\,eV). The agreement between the results for the mixed films and the modelled spectra is almost perfect for all photon energies. Only for the 1:1 blend there seems to be a small deviation around a binding energy of 3\,eV, which is quite sensitive to the assumed mixing ratio. The CuPC shifts in the 1:1 mixture are slightly larger (-360$\pm$35 meV) than in the 10:1 blend (-240$\pm$60\,meV), while the shifts in the C$_{60}$ component are slightly smaller for the 1:1 blend (150$\pm$50\,meV) compared to 260$\pm$100\,meV for the 10:1 blend. It is difficult to judge, whether this difference is significant, since the uncertainty for the contribution of the C$_{60}$ spectral features to the overall spectrum of the 10:1 blend is quite large. It is remarkable however that the relative alignment of the two constituents of this bulk heterojunction remains identical at about 500\,meV shift measured relative to the electronic states of the separate systems. This important result shows that the energetics of the exciton dissociation process are identical, independent of the mixing ratio.

An additional important feature is, that there is no significant broadening of the spectral contributions observed. This indicates the lack of direct hybridization of the orbitals, which is also consistent with earlier observations.\cite{Opitz2012,Molodtsova2006} 

In general, the HOMO and the total spectrum of the CuPC shifts to lower binding energies in the blends, while C$_{60}$ shifts to higher binding energies. This is comparable to the observations made for thin films of CuPC and C$_{60}$ in a layer by layer deposition.\cite{Molodtsova2006} For layered systems this is attributed to an interface dipole,\cite{Grobosch2009,Sai2012,Ren2012} which amounts to 0.5\,eV for the CuPC:C$_{60}$ system independent of the sequence of deposition. We also find a total value of 0.5\,eV for the interface dipole, for both mixing ratios. Since this interface dipole is formed already at monolayer coverage, such an interface dipole formation may well be applicable to  amorphous or nanocrystalline CuPC domains covered at the interfaces with C$_{60}$.\cite{Sai2012} That's why both heterojunctions exhibit an analogous electronic behavior, independent of the C$_{60}$ concentration. This also solves the puzzle that the exciton dissociation does not vary significantly with the ratio of the admixture of C$_{60}$ to the films.\cite{Schuenemann2012}

\section{Conclusion}

We have studied mixtures (blends) of CuPC and C$_{60}$ as model systems for bulk heterojunction organic materials used in photovoltaic applications. For a 1:1 mixing ratio a spontaneous phase separation into domains of tens of nm in size is observed which has similar electronic properties as a layered system. An electronic interface dipole of about 0.5\,eV is observed to be established between the two phases, which facilitates exciton dissociation into separated charge carriers.  Similar electronic behavior is also observed for low admixtures of C$_{60}$. The LEEM images of these films are suggestive of a network of (electron) conductive pathways to be formed, attributed here to C$_{60}$ decorating the boundaries of the amorphous or microcrystalline CuPC grains in the film. This explains the puzzling question why there is only a small difference in performance in photovoltaic applications for heterojunctions over very wide ranges of C$_{60}$ content.

\begin{acknowledgments}
The photoemission studies were performed at the Metrology Lights Source (MLS) of the Physikalisch Technische Bundesanstalt (PTB) in Berlin. We would like to thank the staff of the MLS and especially H. Kaser and T. Reichel  for their experimental support.
\end{acknowledgments}


\begin{thebibliography}{29}%
\makeatletter
\providecommand \@ifxundefined [1]{%
 \@ifx{#1\undefined}
}%
\providecommand \@ifnum [1]{%
 \ifnum #1\expandafter \@firstoftwo
 \else \expandafter \@secondoftwo
 \fi
}%
\providecommand \@ifx [1]{%
 \ifx #1\expandafter \@firstoftwo
 \else \expandafter \@secondoftwo
 \fi
}%
\providecommand \natexlab [1]{#1}%
\providecommand \enquote  [1]{``#1''}%
\providecommand \bibnamefont  [1]{#1}%
\providecommand \bibfnamefont [1]{#1}%
\providecommand \citenamefont [1]{#1}%
\providecommand \href@noop [0]{\@secondoftwo}%
\providecommand \href [0]{\begingroup \@sanitize@url \@href}%
\providecommand \@href[1]{\@@startlink{#1}\@@href}%
\providecommand \@@href[1]{\endgroup#1\@@endlink}%
\providecommand \@sanitize@url [0]{\catcode `\\12\catcode `\$12\catcode
  `\&12\catcode `\#12\catcode `\^12\catcode `\_12\catcode `\%12\relax}%
\providecommand \@@startlink[1]{}%
\providecommand \@@endlink[0]{}%
\providecommand \url  [0]{\begingroup\@sanitize@url \@url }%
\providecommand \@url [1]{\endgroup\@href {#1}{\urlprefix }}%
\providecommand \urlprefix  [0]{URL }%
\providecommand \Eprint [0]{\href }%
\providecommand \doibase [0]{http://dx.doi.org/}%
\providecommand \selectlanguage [0]{\@gobble}%
\providecommand \bibinfo  [0]{\@secondoftwo}%
\providecommand \bibfield  [0]{\@secondoftwo}%
\providecommand \translation [1]{[#1]}%
\providecommand \BibitemOpen [0]{}%
\providecommand \bibitemStop [0]{}%
\providecommand \bibitemNoStop [0]{.\EOS\space}%
\providecommand \EOS [0]{\spacefactor3000\relax}%
\providecommand \BibitemShut  [1]{\csname bibitem#1\endcsname}%
\let\auto@bib@innerbib\@empty
\bibitem [{\citenamefont {Crone}\ \emph {et~al.}(2000)\citenamefont {Crone},
  \citenamefont {Dodabalapur}, \citenamefont {Lin}, \citenamefont {Filas},
  \citenamefont {Bao}, \citenamefont {LaDuca}, \citenamefont {Sarpeshkar},
  \citenamefont {Katz},\ and\ \citenamefont {Li}}]{Crone2000}%
  \BibitemOpen
  \bibfield  {author} {\bibinfo {author} {\bibfnamefont {B.}~\bibnamefont
  {Crone}}, \bibinfo {author} {\bibfnamefont {A.}~\bibnamefont {Dodabalapur}},
  \bibinfo {author} {\bibfnamefont {Y.~Y.}\ \bibnamefont {Lin}}, \bibinfo
  {author} {\bibfnamefont {R.~W.}\ \bibnamefont {Filas}}, \bibinfo {author}
  {\bibfnamefont {Z.}~\bibnamefont {Bao}}, \bibinfo {author} {\bibfnamefont
  {A.}~\bibnamefont {LaDuca}}, \bibinfo {author} {\bibfnamefont
  {R.}~\bibnamefont {Sarpeshkar}}, \bibinfo {author} {\bibfnamefont {H.~E.}\
  \bibnamefont {Katz}}, \ and\ \bibinfo {author} {\bibfnamefont
  {W.}~\bibnamefont {Li}},\ }\href {http://dx.doi.org/10.1038/35000530}
  {\bibfield  {journal} {\bibinfo  {journal} {Nature}\ }\textbf {\bibinfo
  {volume} {403}},\ \bibinfo {pages} {521} (\bibinfo {year}
  {2000})}\BibitemShut {NoStop}%
\bibitem [{\citenamefont {Derycke}\ \emph {et~al.}(2001)\citenamefont
  {Derycke}, \citenamefont {Martel}, \citenamefont {Appenzeller},\ and\
  \citenamefont {Avouris}}]{Derycke2001}%
  \BibitemOpen
  \bibfield  {author} {\bibinfo {author} {\bibfnamefont {V.}~\bibnamefont
  {Derycke}}, \bibinfo {author} {\bibfnamefont {R.}~\bibnamefont {Martel}},
  \bibinfo {author} {\bibfnamefont {J.}~\bibnamefont {Appenzeller}}, \ and\
  \bibinfo {author} {\bibfnamefont {P.}~\bibnamefont {Avouris}},\ }\href
  {\doibase 10.1021/nl015606f} {\bibfield  {journal} {\bibinfo  {journal} {Nano
  Lett.}\ }\textbf {\bibinfo {volume} {1}},\ \bibinfo {pages} {453} (\bibinfo
  {year} {2001})}\BibitemShut {NoStop}%
\bibitem [{\citenamefont {Credi}\ \emph {et~al.}(1997)\citenamefont {Credi},
  \citenamefont {Balzani}, \citenamefont {Langford},\ and\ \citenamefont
  {Stoddart}}]{Credi1997}%
  \BibitemOpen
  \bibfield  {author} {\bibinfo {author} {\bibfnamefont {A.}~\bibnamefont
  {Credi}}, \bibinfo {author} {\bibfnamefont {V.}~\bibnamefont {Balzani}},
  \bibinfo {author} {\bibfnamefont {S.~J.}\ \bibnamefont {Langford}}, \ and\
  \bibinfo {author} {\bibfnamefont {J.~F.}\ \bibnamefont {Stoddart}},\ }\href
  {\doibase 10.1021/ja963572l} {\bibfield  {journal} {\bibinfo  {journal} {J.
  Am. Chem. Soc.}\ }\textbf {\bibinfo {volume} {119}},\ \bibinfo {pages} {2679}
  (\bibinfo {year} {1997})}\BibitemShut {NoStop}%
\bibitem [{\citenamefont {Magri}\ \emph {et~al.}(2006)\citenamefont {Magri},
  \citenamefont {Brown}, \citenamefont {McClean},\ and\ \citenamefont
  {de~Silva}}]{Magri2006}%
  \BibitemOpen
  \bibfield  {author} {\bibinfo {author} {\bibfnamefont {D.~C.}\ \bibnamefont
  {Magri}}, \bibinfo {author} {\bibfnamefont {G.~J.}\ \bibnamefont {Brown}},
  \bibinfo {author} {\bibfnamefont {G.~D.}\ \bibnamefont {McClean}}, \ and\
  \bibinfo {author} {\bibfnamefont {A.~P.}\ \bibnamefont {de~Silva}},\ }\href
  {\doibase 10.1021/ja058295+} {\bibfield  {journal} {\bibinfo  {journal} {J.
  Am. Chem. Soc.}\ }\textbf {\bibinfo {volume} {128}},\ \bibinfo {pages} {4950}
  (\bibinfo {year} {2006})}\BibitemShut {NoStop}%
\bibitem [{\citenamefont {Friend}\ \emph {et~al.}(1999)\citenamefont {Friend},
  \citenamefont {Gymer}, \citenamefont {Holmes}, \citenamefont {Burroughes},
  \citenamefont {Marks}, \citenamefont {Taliani}, \citenamefont {Bradley},
  \citenamefont {Santos}, \citenamefont {Bredas}, \citenamefont {Logdlund},\
  and\ \citenamefont {Salaneck}}]{Friend1999}%
  \BibitemOpen
  \bibfield  {author} {\bibinfo {author} {\bibfnamefont {R.~H.}\ \bibnamefont
  {Friend}}, \bibinfo {author} {\bibfnamefont {R.~W.}\ \bibnamefont {Gymer}},
  \bibinfo {author} {\bibfnamefont {A.~B.}\ \bibnamefont {Holmes}}, \bibinfo
  {author} {\bibfnamefont {J.~H.}\ \bibnamefont {Burroughes}}, \bibinfo
  {author} {\bibfnamefont {R.~N.}\ \bibnamefont {Marks}}, \bibinfo {author}
  {\bibfnamefont {C.}~\bibnamefont {Taliani}}, \bibinfo {author} {\bibfnamefont
  {D.~D.~C.}\ \bibnamefont {Bradley}}, \bibinfo {author} {\bibfnamefont
  {D.~A.~D.}\ \bibnamefont {Santos}}, \bibinfo {author} {\bibfnamefont {J.~L.}\
  \bibnamefont {Bredas}}, \bibinfo {author} {\bibfnamefont {M.}~\bibnamefont
  {Logdlund}}, \ and\ \bibinfo {author} {\bibfnamefont {W.~R.}\ \bibnamefont
  {Salaneck}},\ }\href {http://dx.doi.org/10.1038/16393} {\bibfield  {journal}
  {\bibinfo  {journal} {Nature}\ }\textbf {\bibinfo {volume} {397}},\ \bibinfo
  {pages} {121} (\bibinfo {year} {1999})}\BibitemShut {NoStop}%
\bibitem [{\citenamefont {Meerheim}\ \emph {et~al.}(2009)\citenamefont
  {Meerheim}, \citenamefont {L\"ussem},\ and\ \citenamefont
  {Leo}}]{Meerheim2009}%
  \BibitemOpen
  \bibfield  {author} {\bibinfo {author} {\bibfnamefont {R.}~\bibnamefont
  {Meerheim}}, \bibinfo {author} {\bibfnamefont {B.}~\bibnamefont {L\"ussem}},
  \ and\ \bibinfo {author} {\bibfnamefont {K.}~\bibnamefont {Leo}},\ }\href
  {\doibase 10.1109/JPROC.2009.2022418} {\bibfield  {journal} {\bibinfo
  {journal} {Proc. IEEE}\ }\textbf {\bibinfo {volume} {97}},\ \bibinfo {pages}
  {1606} (\bibinfo {year} {2009})}\BibitemShut {NoStop}%
\bibitem [{\citenamefont {Weichsel}\ \emph {et~al.}(2012)\citenamefont
  {Weichsel}, \citenamefont {Reineke}, \citenamefont {Furno}, \citenamefont
  {L\"ussem},\ and\ \citenamefont {Leo}}]{Weichsel2012}%
  \BibitemOpen
  \bibfield  {author} {\bibinfo {author} {\bibfnamefont {C.}~\bibnamefont
  {Weichsel}}, \bibinfo {author} {\bibfnamefont {S.}~\bibnamefont {Reineke}},
  \bibinfo {author} {\bibfnamefont {M.}~\bibnamefont {Furno}}, \bibinfo
  {author} {\bibfnamefont {B.}~\bibnamefont {L\"ussem}}, \ and\ \bibinfo
  {author} {\bibfnamefont {K.}~\bibnamefont {Leo}},\ }\href {\doibase
  http://dx.doi.org/10.1063/1.3679549} {\bibfield  {journal} {\bibinfo
  {journal} {J. Appl. Phys.}\ }\textbf {\bibinfo {volume} {111}},\ \bibinfo
  {eid} {033102} (\bibinfo {year} {2012})}\BibitemShut {NoStop}%
\bibitem [{\citenamefont {Riede}\ \emph {et~al.}(2008)\citenamefont {Riede},
  \citenamefont {Mueller}, \citenamefont {Tress}, \citenamefont {Schueppel},\
  and\ \citenamefont {Leo}}]{Leo2008}%
  \BibitemOpen
  \bibfield  {author} {\bibinfo {author} {\bibfnamefont {M.}~\bibnamefont
  {Riede}}, \bibinfo {author} {\bibfnamefont {T.}~\bibnamefont {Mueller}},
  \bibinfo {author} {\bibfnamefont {W.}~\bibnamefont {Tress}}, \bibinfo
  {author} {\bibfnamefont {R.}~\bibnamefont {Schueppel}}, \ and\ \bibinfo
  {author} {\bibfnamefont {K.}~\bibnamefont {Leo}},\ }\href
  {http://stacks.iop.org/0957-4484/19/i=42/a=424001} {\bibfield  {journal}
  {\bibinfo  {journal} {Nanotechnology}\ }\textbf {\bibinfo {volume} {19}},\
  \bibinfo {pages} {424001} (\bibinfo {year} {2008})}\BibitemShut {NoStop}%
\bibitem [{\citenamefont {Himpsel}\ \emph {et~al.}(2012)\citenamefont
  {Himpsel}, \citenamefont {Cook}, \citenamefont {de~la Torre}, \citenamefont
  {Garcia-Lastra}, \citenamefont {Gonzalez-Moreno}, \citenamefont {Guo},
  \citenamefont {Hamers}, \citenamefont {Kronawitter}, \citenamefont {Johnson},
  \citenamefont {Ortega}, \citenamefont {Pickup}, \citenamefont {Ragoussi},
  \citenamefont {Rogero}, \citenamefont {Rubio}, \citenamefont {Ruther},
  \citenamefont {Vayssieres}, \citenamefont {Yang},\ and\ \citenamefont
  {Zegkinoglou}}]{Himpsel2012}%
  \BibitemOpen
  \bibfield  {author} {\bibinfo {author} {\bibfnamefont {F.}~\bibnamefont
  {Himpsel}}, \bibinfo {author} {\bibfnamefont {P.}~\bibnamefont {Cook}},
  \bibinfo {author} {\bibfnamefont {G.}~\bibnamefont {de~la Torre}}, \bibinfo
  {author} {\bibfnamefont {J.}~\bibnamefont {Garcia-Lastra}}, \bibinfo {author}
  {\bibfnamefont {R.}~\bibnamefont {Gonzalez-Moreno}}, \bibinfo {author}
  {\bibfnamefont {J.-H.}\ \bibnamefont {Guo}}, \bibinfo {author} {\bibfnamefont
  {R.}~\bibnamefont {Hamers}}, \bibinfo {author} {\bibfnamefont
  {C.}~\bibnamefont {Kronawitter}}, \bibinfo {author} {\bibfnamefont
  {P.}~\bibnamefont {Johnson}}, \bibinfo {author} {\bibfnamefont
  {J.}~\bibnamefont {Ortega}}, \bibinfo {author} {\bibfnamefont
  {D.}~\bibnamefont {Pickup}}, \bibinfo {author} {\bibfnamefont {M.-E.}\
  \bibnamefont {Ragoussi}}, \bibinfo {author} {\bibfnamefont {C.}~\bibnamefont
  {Rogero}}, \bibinfo {author} {\bibfnamefont {A.}~\bibnamefont {Rubio}},
  \bibinfo {author} {\bibfnamefont {R.}~\bibnamefont {Ruther}}, \bibinfo
  {author} {\bibfnamefont {L.}~\bibnamefont {Vayssieres}}, \bibinfo {author}
  {\bibfnamefont {W.}~\bibnamefont {Yang}}, \ and\ \bibinfo {author}
  {\bibfnamefont {I.}~\bibnamefont {Zegkinoglou}},\ }\href {\doibase
  http://dx.doi.org/10.1016/j.elspec.2012.10.002} {\bibfield  {journal}
  {\bibinfo  {journal} {J. Electron. Spectrosc. Relat. Phenom.}\ } (\bibinfo
  {year} {2012}),\ http://dx.doi.org/10.1016/j.elspec.2012.10.002}\BibitemShut
  {NoStop}%
\bibitem [{\citenamefont {Pickup}\ \emph {et~al.}(2013)\citenamefont {Pickup},
  \citenamefont {Zegkinoglou}, \citenamefont {Ballesteros}, \citenamefont
  {Ganivet}, \citenamefont {Garc\'ia-Lastra}, \citenamefont {Cook},
  \citenamefont {Johnson}, \citenamefont {Rogero}, \citenamefont {de~Groot},
  \citenamefont {Rubio}, \citenamefont {de~la Torre}, \citenamefont {Ortega},\
  and\ \citenamefont {Himpsel}}]{Pickup2013}%
  \BibitemOpen
  \bibfield  {author} {\bibinfo {author} {\bibfnamefont {D.~F.}\ \bibnamefont
  {Pickup}}, \bibinfo {author} {\bibfnamefont {I.}~\bibnamefont {Zegkinoglou}},
  \bibinfo {author} {\bibfnamefont {B.}~\bibnamefont {Ballesteros}}, \bibinfo
  {author} {\bibfnamefont {C.~R.}\ \bibnamefont {Ganivet}}, \bibinfo {author}
  {\bibfnamefont {J.~M.}\ \bibnamefont {Garc\'ia-Lastra}}, \bibinfo {author}
  {\bibfnamefont {P.~L.}\ \bibnamefont {Cook}}, \bibinfo {author}
  {\bibfnamefont {P.~S.}\ \bibnamefont {Johnson}}, \bibinfo {author}
  {\bibfnamefont {C.}~\bibnamefont {Rogero}}, \bibinfo {author} {\bibfnamefont
  {F.}~\bibnamefont {de~Groot}}, \bibinfo {author} {\bibfnamefont
  {A.}~\bibnamefont {Rubio}}, \bibinfo {author} {\bibfnamefont
  {G.}~\bibnamefont {de~la Torre}}, \bibinfo {author} {\bibfnamefont {J.~E.}\
  \bibnamefont {Ortega}}, \ and\ \bibinfo {author} {\bibfnamefont {F.~J.}\
  \bibnamefont {Himpsel}},\ }\href {\doibase 10.1021/jp3117853} {\bibfield
  {journal} {\bibinfo  {journal} {J. Phys. Chem. C}\ }\textbf {\bibinfo
  {volume} {117}},\ \bibinfo {pages} {4410} (\bibinfo {year}
  {2013})}\BibitemShut {NoStop}%
\bibitem [{\citenamefont {Tang}(1986)}]{Tang1986}%
  \BibitemOpen
  \bibfield  {author} {\bibinfo {author} {\bibfnamefont {C.~W.}\ \bibnamefont
  {Tang}},\ }\href {\doibase http://dx.doi.org/10.1063/1.96937} {\bibfield
  {journal} {\bibinfo  {journal} {Appl. Phys. Lett.}\ }\textbf {\bibinfo
  {volume} {48}},\ \bibinfo {pages} {183} (\bibinfo {year} {1986})}\BibitemShut
  {NoStop}%
\bibitem [{\citenamefont {Sariciftci}\ \emph {et~al.}(1992)\citenamefont
  {Sariciftci}, \citenamefont {Smilowitz}, \citenamefont {Heeger},\ and\
  \citenamefont {Wudl}}]{Sariciftci1992}%
  \BibitemOpen
  \bibfield  {author} {\bibinfo {author} {\bibfnamefont {N.~S.}\ \bibnamefont
  {Sariciftci}}, \bibinfo {author} {\bibfnamefont {L.}~\bibnamefont
  {Smilowitz}}, \bibinfo {author} {\bibfnamefont {A.~J.}\ \bibnamefont
  {Heeger}}, \ and\ \bibinfo {author} {\bibfnamefont {F.}~\bibnamefont
  {Wudl}},\ }\href {\doibase 10.1126/science.258.5087.1474} {\bibfield
  {journal} {\bibinfo  {journal} {Science}\ }\textbf {\bibinfo {volume}
  {258}},\ \bibinfo {pages} {1474} (\bibinfo {year} {1992})}\BibitemShut
  {NoStop}%
\bibitem [{\citenamefont {Gunnarsson}\ \emph {et~al.}(1995)\citenamefont
  {Gunnarsson}, \citenamefont {Handschuh}, \citenamefont {Bechthold},
  \citenamefont {Kessler}, \citenamefont {Gantef\"or},\ and\ \citenamefont
  {Eberhardt}}]{Gunnarsson1995}%
  \BibitemOpen
  \bibfield  {author} {\bibinfo {author} {\bibfnamefont {O.}~\bibnamefont
  {Gunnarsson}}, \bibinfo {author} {\bibfnamefont {H.}~\bibnamefont
  {Handschuh}}, \bibinfo {author} {\bibfnamefont {P.~S.}\ \bibnamefont
  {Bechthold}}, \bibinfo {author} {\bibfnamefont {B.}~\bibnamefont {Kessler}},
  \bibinfo {author} {\bibfnamefont {G.}~\bibnamefont {Gantef\"or}}, \ and\
  \bibinfo {author} {\bibfnamefont {W.}~\bibnamefont {Eberhardt}},\ }\href
  {\doibase 10.1103/PhysRevLett.74.1875} {\bibfield  {journal} {\bibinfo
  {journal} {Phys. Rev. Lett.}\ }\textbf {\bibinfo {volume} {74}},\ \bibinfo
  {pages} {1875} (\bibinfo {year} {1995})}\BibitemShut {NoStop}%
\bibitem [{\citenamefont {Schlebusch}\ \emph {et~al.}(1996)\citenamefont
  {Schlebusch}, \citenamefont {Kessler}, \citenamefont {Cramm},\ and\
  \citenamefont {Eberhardt}}]{Schlebusch1996}%
  \BibitemOpen
  \bibfield  {author} {\bibinfo {author} {\bibfnamefont {C.}~\bibnamefont
  {Schlebusch}}, \bibinfo {author} {\bibfnamefont {B.}~\bibnamefont {Kessler}},
  \bibinfo {author} {\bibfnamefont {S.}~\bibnamefont {Cramm}}, \ and\ \bibinfo
  {author} {\bibfnamefont {W.}~\bibnamefont {Eberhardt}},\ }\href {\doibase
  http://dx.doi.org/10.1016/0379-6779(96)80077-4} {\bibfield  {journal}
  {\bibinfo  {journal} {Synth. Met.}\ }\textbf {\bibinfo {volume} {77}},\
  \bibinfo {pages} {151 } (\bibinfo {year} {1996})}\BibitemShut {NoStop}%
\bibitem [{\citenamefont {Schlebusch}\ \emph {et~al.}(1999)\citenamefont
  {Schlebusch}, \citenamefont {Morenzin}, \citenamefont {Kessler},\ and\
  \citenamefont {Eberhardt}}]{Schlebusch1999}%
  \BibitemOpen
  \bibfield  {author} {\bibinfo {author} {\bibfnamefont {C.}~\bibnamefont
  {Schlebusch}}, \bibinfo {author} {\bibfnamefont {J.}~\bibnamefont
  {Morenzin}}, \bibinfo {author} {\bibfnamefont {B.}~\bibnamefont {Kessler}}, \
  and\ \bibinfo {author} {\bibfnamefont {W.}~\bibnamefont {Eberhardt}},\ }\href
  {\doibase http://dx.doi.org/10.1016/S0008-6223(98)00260-7} {\bibfield
  {journal} {\bibinfo  {journal} {Carbon}\ }\textbf {\bibinfo {volume} {37}},\
  \bibinfo {pages} {717 } (\bibinfo {year} {1999})}\BibitemShut {NoStop}%
\bibitem [{\citenamefont {Kessler}\ \emph {et~al.}(1998)\citenamefont
  {Kessler}, \citenamefont {Schlebusch}, \citenamefont {Morenzin},\ and\
  \citenamefont {Eberhardt}}]{Kessler1998}%
  \BibitemOpen
  \bibfield  {author} {\bibinfo {author} {\bibfnamefont {B.}~\bibnamefont
  {Kessler}}, \bibinfo {author} {\bibfnamefont {C.}~\bibnamefont {Schlebusch}},
  \bibinfo {author} {\bibfnamefont {J.}~\bibnamefont {Morenzin}}, \ and\
  \bibinfo {author} {\bibfnamefont {W.}~\bibnamefont {Eberhardt}},\ }\href@noop
  {} {\bibfield  {journal} {\bibinfo  {journal} {AIP Conf. Proc.}\ }\textbf
  {\bibinfo {volume} {442}} (\bibinfo {year} {1998})}\BibitemShut {NoStop}%
\bibitem [{\citenamefont {Fahlman}\ \emph {et~al.}(2013)\citenamefont
  {Fahlman}, \citenamefont {Sehati}, \citenamefont {Osikowicz}, \citenamefont
  {Braun}, \citenamefont {de~Jong},\ and\ \citenamefont
  {Brocks}}]{Fahlman2013}%
  \BibitemOpen
  \bibfield  {author} {\bibinfo {author} {\bibfnamefont {M.}~\bibnamefont
  {Fahlman}}, \bibinfo {author} {\bibfnamefont {P.}~\bibnamefont {Sehati}},
  \bibinfo {author} {\bibfnamefont {W.}~\bibnamefont {Osikowicz}}, \bibinfo
  {author} {\bibfnamefont {S.}~\bibnamefont {Braun}}, \bibinfo {author}
  {\bibfnamefont {M.~P.}\ \bibnamefont {de~Jong}}, \ and\ \bibinfo {author}
  {\bibfnamefont {G.}~\bibnamefont {Brocks}},\ }\href {\doibase
  http://dx.doi.org/10.1016/j.elspec.2013.02.001} {\bibfield  {journal}
  {\bibinfo  {journal} {J. Electron. Spectrosc. Relat. Phenom.}\ } (\bibinfo
  {year} {2013}),\ http://dx.doi.org/10.1016/j.elspec.2013.02.001},\ \bibinfo
  {note} {and references therein}\BibitemShut {NoStop}%
\bibitem [{\citenamefont {Opitz}\ \emph {et~al.}(2012)\citenamefont {Opitz},
  \citenamefont {Frisch}, \citenamefont {Schlesinger}, \citenamefont {Wilke},\
  and\ \citenamefont {Koch}}]{Opitz2012}%
  \BibitemOpen
  \bibfield  {author} {\bibinfo {author} {\bibfnamefont {A.}~\bibnamefont
  {Opitz}}, \bibinfo {author} {\bibfnamefont {J.}~\bibnamefont {Frisch}},
  \bibinfo {author} {\bibfnamefont {R.}~\bibnamefont {Schlesinger}}, \bibinfo
  {author} {\bibfnamefont {A.}~\bibnamefont {Wilke}}, \ and\ \bibinfo {author}
  {\bibfnamefont {N.}~\bibnamefont {Koch}},\ }\href {\doibase
  http://dx.doi.org/10.1016/j.elspec.2012.11.008} {\bibfield  {journal}
  {\bibinfo  {journal} {J. Electron. Spectrosc. Relat. Phenom.}\ } (\bibinfo
  {year} {2012}),\ http://dx.doi.org/10.1016/j.elspec.2012.11.008},\ \bibinfo
  {note} {and references therein}\BibitemShut {NoStop}%
\bibitem [{\citenamefont {Heremans}\ \emph {et~al.}(2009)\citenamefont
  {Heremans}, \citenamefont {Cheyns},\ and\ \citenamefont
  {Rand}}]{Heremans2009}%
  \BibitemOpen
  \bibfield  {author} {\bibinfo {author} {\bibfnamefont {P.}~\bibnamefont
  {Heremans}}, \bibinfo {author} {\bibfnamefont {D.}~\bibnamefont {Cheyns}}, \
  and\ \bibinfo {author} {\bibfnamefont {B.~P.}\ \bibnamefont {Rand}},\ }\href
  {\doibase 10.1021/ar9000923} {\bibfield  {journal} {\bibinfo  {journal} {Acc.
  Chem. Res.}\ }\textbf {\bibinfo {volume} {42}},\ \bibinfo {pages} {1740}
  (\bibinfo {year} {2009})}\BibitemShut {NoStop}%
\bibitem [{\citenamefont {Sch\"unemann}\ \emph {et~al.}(2012)\citenamefont
  {Sch\"unemann}, \citenamefont {Wynands}, \citenamefont {Wilde}, \citenamefont
  {Hein}, \citenamefont {Pf\"utzner}, \citenamefont {Elschner}, \citenamefont
  {Eichhorn}, \citenamefont {Leo},\ and\ \citenamefont
  {Riede}}]{Schuenemann2012}%
  \BibitemOpen
  \bibfield  {author} {\bibinfo {author} {\bibfnamefont {C.}~\bibnamefont
  {Sch\"unemann}}, \bibinfo {author} {\bibfnamefont {D.}~\bibnamefont
  {Wynands}}, \bibinfo {author} {\bibfnamefont {L.}~\bibnamefont {Wilde}},
  \bibinfo {author} {\bibfnamefont {M.~P.}\ \bibnamefont {Hein}}, \bibinfo
  {author} {\bibfnamefont {S.}~\bibnamefont {Pf\"utzner}}, \bibinfo {author}
  {\bibfnamefont {C.}~\bibnamefont {Elschner}}, \bibinfo {author}
  {\bibfnamefont {K.-J.}\ \bibnamefont {Eichhorn}}, \bibinfo {author}
  {\bibfnamefont {K.}~\bibnamefont {Leo}}, \ and\ \bibinfo {author}
  {\bibfnamefont {M.}~\bibnamefont {Riede}},\ }\href {\doibase
  10.1103/PhysRevB.85.245314} {\bibfield  {journal} {\bibinfo  {journal} {Phys.
  Rev. B}\ }\textbf {\bibinfo {volume} {85}},\ \bibinfo {pages} {245314}
  (\bibinfo {year} {2012})}\BibitemShut {NoStop}%
\bibitem [{\citenamefont {Kim}\ \emph {et~al.}(2011)\citenamefont {Kim},
  \citenamefont {Kim}, \citenamefont {Lee}, \citenamefont {Kim}, \citenamefont
  {Jang},\ and\ \citenamefont {Kim}}]{Kim2011}%
  \BibitemOpen
  \bibfield  {author} {\bibinfo {author} {\bibfnamefont {H.~J.}\ \bibnamefont
  {Kim}}, \bibinfo {author} {\bibfnamefont {J.~W.}\ \bibnamefont {Kim}},
  \bibinfo {author} {\bibfnamefont {H.~H.}\ \bibnamefont {Lee}}, \bibinfo
  {author} {\bibfnamefont {T.-M.}\ \bibnamefont {Kim}}, \bibinfo {author}
  {\bibfnamefont {J.}~\bibnamefont {Jang}}, \ and\ \bibinfo {author}
  {\bibfnamefont {J.-J.}\ \bibnamefont {Kim}},\ }\href {\doibase
  10.1021/jz200724x} {\bibfield  {journal} {\bibinfo  {journal} {J. Phys. Chem.
  Lett.}\ }\textbf {\bibinfo {volume} {2}},\ \bibinfo {pages} {1710} (\bibinfo
  {year} {2011})}\BibitemShut {NoStop}%
\bibitem [{\citenamefont {Lupulescu}\ \emph {et~al.}(2013)\citenamefont
  {Lupulescu}, \citenamefont {Arion}, \citenamefont {Hergenhahn}, \citenamefont
  {Ovsyannikov}, \citenamefont {F\"orstel}, \citenamefont {Gavrila},\ and\
  \citenamefont {Eberhardt}}]{Lupulescu2013}%
  \BibitemOpen
  \bibfield  {author} {\bibinfo {author} {\bibfnamefont {C.}~\bibnamefont
  {Lupulescu}}, \bibinfo {author} {\bibfnamefont {T.}~\bibnamefont {Arion}},
  \bibinfo {author} {\bibfnamefont {U.}~\bibnamefont {Hergenhahn}}, \bibinfo
  {author} {\bibfnamefont {R.}~\bibnamefont {Ovsyannikov}}, \bibinfo {author}
  {\bibfnamefont {M.}~\bibnamefont {F\"orstel}}, \bibinfo {author}
  {\bibfnamefont {G.}~\bibnamefont {Gavrila}}, \ and\ \bibinfo {author}
  {\bibfnamefont {W.}~\bibnamefont {Eberhardt}},\ }\href {\doibase
  http://dx.doi.org/10.1016/j.elspec.2013.09.002} {\bibfield  {journal}
  {\bibinfo  {journal} {J. Electron. Spectrosc. Relat. Phenom.}\ } (\bibinfo
  {year} {2013}),\ http://dx.doi.org/10.1016/j.elspec.2013.09.002}\BibitemShut
  {NoStop}%
\bibitem [{\citenamefont {Gottwald}\ \emph {et~al.}(2012)\citenamefont
  {Gottwald}, \citenamefont {Klein}, \citenamefont {M\"uller}, \citenamefont
  {Richter}, \citenamefont {Scholze}, \citenamefont {Thornagel},\ and\
  \citenamefont {Ulm}}]{Gottwald2012}%
  \BibitemOpen
  \bibfield  {author} {\bibinfo {author} {\bibfnamefont {A.}~\bibnamefont
  {Gottwald}}, \bibinfo {author} {\bibfnamefont {R.}~\bibnamefont {Klein}},
  \bibinfo {author} {\bibfnamefont {R.}~\bibnamefont {M\"uller}}, \bibinfo
  {author} {\bibfnamefont {M.}~\bibnamefont {Richter}}, \bibinfo {author}
  {\bibfnamefont {F.}~\bibnamefont {Scholze}}, \bibinfo {author} {\bibfnamefont
  {R.}~\bibnamefont {Thornagel}}, \ and\ \bibinfo {author} {\bibfnamefont
  {G.}~\bibnamefont {Ulm}},\ }\href
  {http://stacks.iop.org/0026-1394/49/i=2/a=S146} {\bibfield  {journal}
  {\bibinfo  {journal} {Metrologia}\ }\textbf {\bibinfo {volume} {49}},\
  \bibinfo {pages} {S146} (\bibinfo {year} {2012})}\BibitemShut {NoStop}%
\bibitem [{\citenamefont {Tromp}\ \emph {et~al.}(2010)\citenamefont {Tromp},
  \citenamefont {Hannon}, \citenamefont {Ellis}, \citenamefont {Wan},
  \citenamefont {Berghaus},\ and\ \citenamefont {Schaff}}]{Tromp2010}%
  \BibitemOpen
  \bibfield  {author} {\bibinfo {author} {\bibfnamefont {R.}~\bibnamefont
  {Tromp}}, \bibinfo {author} {\bibfnamefont {J.}~\bibnamefont {Hannon}},
  \bibinfo {author} {\bibfnamefont {A.}~\bibnamefont {Ellis}}, \bibinfo
  {author} {\bibfnamefont {W.}~\bibnamefont {Wan}}, \bibinfo {author}
  {\bibfnamefont {A.}~\bibnamefont {Berghaus}}, \ and\ \bibinfo {author}
  {\bibfnamefont {O.}~\bibnamefont {Schaff}},\ }\href {\doibase
  http://dx.doi.org/10.1016/j.ultramic.2010.03.005} {\bibfield  {journal}
  {\bibinfo  {journal} {Ultramicroscopy}\ }\textbf {\bibinfo {volume} {110}},\
  \bibinfo {pages} {852 } (\bibinfo {year} {2010})}\BibitemShut {NoStop}%
\bibitem [{\citenamefont {Tromp}\ \emph {et~al.}(2009)\citenamefont {Tromp},
  \citenamefont {Fujikawa}, \citenamefont {Hannon}, \citenamefont {Ellis},
  \citenamefont {Berghaus},\ and\ \citenamefont {Schaff}}]{Tromp2009}%
  \BibitemOpen
  \bibfield  {author} {\bibinfo {author} {\bibfnamefont {R.~M.}\ \bibnamefont
  {Tromp}}, \bibinfo {author} {\bibfnamefont {Y.}~\bibnamefont {Fujikawa}},
  \bibinfo {author} {\bibfnamefont {J.~B.}\ \bibnamefont {Hannon}}, \bibinfo
  {author} {\bibfnamefont {A.~W.}\ \bibnamefont {Ellis}}, \bibinfo {author}
  {\bibfnamefont {A.}~\bibnamefont {Berghaus}}, \ and\ \bibinfo {author}
  {\bibfnamefont {O.}~\bibnamefont {Schaff}},\ }\href
  {http://stacks.iop.org/0953-8984/21/i=31/a=314007} {\bibfield  {journal}
  {\bibinfo  {journal} {J. Phys.: Cond. Matt.}\ }\textbf {\bibinfo {volume}
  {21}},\ \bibinfo {pages} {314007} (\bibinfo {year} {2009})}\BibitemShut
  {NoStop}%
\bibitem [{\citenamefont {Molodtsova}\ and\ \citenamefont
  {Knupfer}(2006)}]{Molodtsova2006}%
  \BibitemOpen
  \bibfield  {author} {\bibinfo {author} {\bibfnamefont {O.~V.}\ \bibnamefont
  {Molodtsova}}\ and\ \bibinfo {author} {\bibfnamefont {M.}~\bibnamefont
  {Knupfer}},\ }\href {\doibase http://dx.doi.org/10.1063/1.2175468} {\bibfield
   {journal} {\bibinfo  {journal} {J. Appl. Phys.}\ }\textbf {\bibinfo {volume}
  {99}},\ \bibinfo {eid} {053704} (\bibinfo {year} {2006})}\BibitemShut
  {NoStop}%
\bibitem [{\citenamefont {Grobosch}\ \emph {et~al.}(2009)\citenamefont
  {Grobosch}, \citenamefont {Aristov}, \citenamefont {Molodtsova},
  \citenamefont {Schmidt}, \citenamefont {Doyle}, \citenamefont {Nannarone},\
  and\ \citenamefont {Knupfer}}]{Grobosch2009}%
  \BibitemOpen
  \bibfield  {author} {\bibinfo {author} {\bibfnamefont {M.}~\bibnamefont
  {Grobosch}}, \bibinfo {author} {\bibfnamefont {V.~Y.}\ \bibnamefont
  {Aristov}}, \bibinfo {author} {\bibfnamefont {O.~V.}\ \bibnamefont
  {Molodtsova}}, \bibinfo {author} {\bibfnamefont {C.}~\bibnamefont {Schmidt}},
  \bibinfo {author} {\bibfnamefont {B.~P.}\ \bibnamefont {Doyle}}, \bibinfo
  {author} {\bibfnamefont {S.}~\bibnamefont {Nannarone}}, \ and\ \bibinfo
  {author} {\bibfnamefont {M.}~\bibnamefont {Knupfer}},\ }\href {\doibase
  10.1021/jp901731y} {\bibfield  {journal} {\bibinfo  {journal} {J. Phys. Chem.
  C}\ }\textbf {\bibinfo {volume} {113}},\ \bibinfo {pages} {13219} (\bibinfo
  {year} {2009})}\BibitemShut {NoStop}%
\bibitem [{\citenamefont {Sai}\ \emph {et~al.}(2012)\citenamefont {Sai},
  \citenamefont {Gearba}, \citenamefont {Dolocan}, \citenamefont {Tritsch},
  \citenamefont {Chan}, \citenamefont {Chelikowsky}, \citenamefont {Leung},\
  and\ \citenamefont {Zhu}}]{Sai2012}%
  \BibitemOpen
  \bibfield  {author} {\bibinfo {author} {\bibfnamefont {N.}~\bibnamefont
  {Sai}}, \bibinfo {author} {\bibfnamefont {R.}~\bibnamefont {Gearba}},
  \bibinfo {author} {\bibfnamefont {A.}~\bibnamefont {Dolocan}}, \bibinfo
  {author} {\bibfnamefont {J.~R.}\ \bibnamefont {Tritsch}}, \bibinfo {author}
  {\bibfnamefont {W.-L.}\ \bibnamefont {Chan}}, \bibinfo {author}
  {\bibfnamefont {J.~R.}\ \bibnamefont {Chelikowsky}}, \bibinfo {author}
  {\bibfnamefont {K.}~\bibnamefont {Leung}}, \ and\ \bibinfo {author}
  {\bibfnamefont {X.}~\bibnamefont {Zhu}},\ }\href {\doibase 10.1021/jz300744r}
  {\bibfield  {journal} {\bibinfo  {journal} {J. Phys. Chem. Lett.}\ }\textbf
  {\bibinfo {volume} {3}},\ \bibinfo {pages} {2173} (\bibinfo {year}
  {2012})}\BibitemShut {NoStop}%
\bibitem [{\citenamefont {Ren}\ \emph {et~al.}(2012)\citenamefont {Ren},
  \citenamefont {Meng},\ and\ \citenamefont {Kaxiras}}]{Ren2012}%
  \BibitemOpen
  \bibfield  {author} {\bibinfo {author} {\bibfnamefont {J.}~\bibnamefont
  {Ren}}, \bibinfo {author} {\bibfnamefont {S.}~\bibnamefont {Meng}}, \ and\
  \bibinfo {author} {\bibfnamefont {E.}~\bibnamefont {Kaxiras}},\ }\href
  {\doibase 10.1007/s12274-012-0204-7} {\bibfield  {journal} {\bibinfo
  {journal} {Nano Research}\ }\textbf {\bibinfo {volume} {5}},\ \bibinfo
  {pages} {248} (\bibinfo {year} {2012})}\BibitemShut {NoStop}%
\end{thebibliography}
\end{document}